# A Diffractive Study of Parametric Process in Nonlinear Photonic Crystals

H. C. Guo[1], Y. Q. Qin[2], G. J. You[1], W. M. Liu[1], S. H. Tang[1]

[1]Physics Department, National University of Singapore, Singapore, 117542

[2] Physics Department, Nanjing University, Nanjing, China

**Abstract**

We report a general description of quasi-phase-matched parametric process in nonlinear photonic crystals (NLPC) by extending the conventional X-ray diffraction theory in solids. Under the virtual wave approximation, phase-matching resonance is equivalent to the diffraction of the scattered virtual wave. Hence a modified NLPC Ewald construction can be built up, which illustrates the nature of the accident for the diffraction of the virtual wave in NLPC and further reveals the complete set of diffractions of the virtual wave for both of the *air-dielectric* and *dielectric-dielectric* contacts. We show the two basic linear sequences, the anti-stacking and para-stacking linear sequences, in one-dimension (1D) NLPC and present a general rule for multiple phase-matching resonances in 1D NLPC. The parameters affecting the NLPC structure factor are investigated, which indicate that not only the Ewald construction but also the relative NLPC atom size together determine whether a diffraction of the virtual wave can occur in 2D NLPC. The results also show that 1D NLPC is a better choice than 2D NLPC for a single parametric process.

The studies of NLPC began with the proposition of the quasi-phase-matching (QPM) concept by Armstrong [1] and Franken [2] independently. QPM refers to the realization of photon momentum conservation i.e., phase-matching, by means of reversing the direction of the nonlinear optical susceptibility after phase slip between the interacting waves accumulates to 180º at every coherence length $l_c$. QPM eliminates the dependence of the realization of phase-matching on the inherent properties of materials and hence has advantages such as non-critical phase matching of any interaction within the transparency range, high nonlinear gain, no walk-off. In 1998, Berger extended the QPM study from 1D to 2D and proposed the concept of NLPC [3]. Since then, NLPCs have been verified to be a valuable platform for the light-matter interaction in nonlinear optical regime due to their unique spatiospectral characteristics that enable them to perform a variety of all-optical functions [4-8].

In a general sight, the coupling of light waves in NLPC and the X-rays scattering in solids share same characteristics: no excitation of the material, inherent wave vector mismatch (in parametric process, wave vector mismatch is from normal dispersion, while in X-rays scattering is from reflection by different atomic layers), and the structure-dependent compensation for the wave vector mismatch. In this Letter, by extending the conventional X-rays diffraction theory in solids, we propose the virtual wave approximation, by which the NLPC Ewald construction is presented that reveals the complete set of phase-matching resonances for both of the *air-dielectric* and *dielectric-dielectric* contacts. We investigate the two basic sequences regarding the stacking of the NLPC primitive cells in 1D NLPC and reveal the mechanism for multiple phase-matching resonances, which is essentially the key for applications such as optical communication networks, coherent laser frequency conversion, and spectroscopy [9-11]. The parameters that affect the NLPC structure factor are investigated, indicating that the Ewald construction and the relative NLPC atom size together determine the diffraction for a virtual wave in 2D NLPC. The results further show that 1D NLPC is a better choice than 2D NLPC for single parametric process.

Analogous to a diatomic crystal, the NLPC primitive cell is defined as the minimum repeating unit which contains a pair of antiparallel ferroelectric domains, as shown in Fig. 1. Accordingly, the domain acts as the equivalent of the atom. Without regard of the energy distribution among the interacting waves, the entire parametric process in a NLPC can be treated as a propagating virtual wave $\xi(\vec{r}) = \xi_0 \exp(-i\vec{K} \cdot \vec{r})$, with wavelength of $\lambda = 2l_c$. $|\xi_0|^2$ carries the total energy of the parametric process and wave vector $\vec{K} = 2\pi/\lambda$ reflects the phase mismatch, which makes the entire energy $|\xi_0|^2$ oscillate between the initial and product waves every coherence length $l_c$, as shown in Fig. 2. As this virtual wave propagates along $\vec{K}$ direction in NLPC, it is also scattered by NLPC structure with the phase factor of $\exp[-i(\vec{K}' - \vec{K}_0)\vec{r}] = \exp[-i(\Delta\vec{K})\vec{r}]$, where $\vec{K}_0$ and $\vec{K}'$ are the wave vectors of the incoming and outgoing virtual waves, and $\Delta\vec{K}$ denotes the scattering vector due to the phase mismatch induced in NLPC.

Similar to X-rays diffraction in solids [12], the total amplitude of the wave scattered from a NLPC in direction $\vec{K}'$ is determined by the Fourier transform of the interaction potential, which is often related to the density of the object that scatters the radiation. In NLPC, it is therefore proportional to the integral over the local distribution $f^{(n)}(\vec{r})$ of nonlinear coefficient $\chi^{(2)}$ times the phase factor $\exp[-i(\Delta\vec{K})\vec{r}]$, which defines the quantity that we call the *NLPC structure factor* $F_{\Delta \vec{K}}^{NLPC}$

$$F_{\Delta \vec{K}}^{NLPC} = \frac{1}{\Pi} \sum_{n=1}^{N} \int d\vec{r} f^{(n)}(\vec{r}) \exp(-i |\Delta \vec{K}| \vec{r}) \qquad (1)$$

$f^{(n)}(\vec{r}) = \Gamma^{(n)}(\vec{r})P$ is accordingly the NLPC *atomic form factor*, where $P$ (i.e., $\chi^{(2)}$) is the equivalent of atomic dipole. The spatial function $\Gamma^{(n)}(\vec{r})$ takes the values of +1 or -1, describing the dipole distribution in a NLPC primitive cell. $N$ and $\Pi$ denote the total primitive cells in NLPC and the entire spatial spreading of the phase-matching topology. According to the diffraction theory, the diffraction of the scattered virtual wave, which is defined as the *photonic Bragg condition,* only occurs if the scattering vector $\Delta \vec{K}$ equals to a NLPC reciprocal lattice (RL) vector $\vec{\Omega}_m$. Phase-matching resonance then occurs and efficient parametric process is achieved.

The above virtual wave approximation enables us to build up the NLPC Ewald construction as shown in Fig. 3, which illustrates the nature of the accident that must occur in order to satisfy the diffraction of the virtual wave in NLPC. In the case of *air-dielectric* contact, $\vec{K}_0 = 0$ and the origin of the RL is the same as the center of the Ewald sphere (C. E.). The possible Bragg conditions are therefore confined in four or eight symmetry-fold in the square Bravais lattice, and six or twelve symmetry-fold in the hexagonal lattice, as shown in Fig. 3(a). For the case of *dielectric-dielectric* contact (Fig. 3(b)), the virtual wave has an initial finite wave vector $\vec{K}_0$ and the photonic Bragg condition is controlled by the following rule: We draw a circle of radius $|\vec{K}_0|$ about the origin $\Gamma$ (the origin of RL). This circle is actually the assembly of C. E. Using any point on the circle as the center, we then draw the Ewald sphere of radius $|\vec{K}'|$. If a point of the RL is located on the Ewald sphere, photonic Bragg condition is envisaged. Finally, all Ewald spheres form an annular zone with the two radius of $|\vec{K}_0|+|\vec{K}'|$ and $||\vec{K}_0|-|\vec{K}'||$, and photonic Bragg diffraction is possible for any RL point in this zone. Comparing Fig. 3(a) with Fig. 3(b), more photonic Bragg diffractions can be induced if the incoming virtual wave has an initial finite wave vector $\vec{K}_0$.

In 1D NLPC, $F_{\vec{\Omega}_m}^{NLPC}$ reduces to a scalar function as $F_{\vec{\Omega}_m}^{NLPC} = \frac{1}{L}\sum_{n=1}^{N}\int dx f^{(n)}(x)\exp(-i|\vec{\Omega}_m|x)$, where $L$ denotes the length along the phase-matching direction. As shown in Fig. 1, depending on whether the neighboring atoms from adjacent NLPC primitive cells have antiparallel or parallel polarization directions, there are two basic linear sequences, the *anti-stacking* sequence and the *para-stacking* sequence, with the following 1D spatial functions of $\Gamma^{(n)}(x) = 1 - 2\text{int}(\frac{x-(n-1)a_0}{a_1}) + 2\text{int}(\frac{x-(n-1)a_0-a_1}{a_1})$ and $\Gamma^{(n)}(x) = [1 - 2\text{int}(\frac{x}{a_0}) + 4\text{int}(\frac{x}{2a_0})][1 - 2\text{int}(\frac{x-(n-1)a_0}{a_1}) + 2\text{int}(\frac{x-(n-1)a_0-a_1}{a_1})]$ respectively, where $\text{int}(x)$ gives the largest integer $\leq x$. Then we have $|F_{\vec{\Omega}_m}^{NLPC}| = 2P\sin(\pi m a_1/a_0)/m\pi$ for anti-stacking sequence, which consists with the conventional QPM theory [13]. The size ratio $a_1/a_0$ between the NLPC atom and primitive cell is defined as the duty cycle $\Phi$, which determines the forbiddance of diffraction by the rule of $ma_1/a_0$=integer. For the para-stacking sequence, we have $|F_{\vec{\Omega}_m}^{NLPC}| = 2P/m\pi$ and 0 for odd $m$ and even $m$ respectively. The forbiddance condition no longer depends on the size ratio $a_1/a_0$ in such sequence. Only diffractions with even order reciprocal vector are forbidden as shown in Fig. 4. Further analysis shows that the additional domain spreading $\delta$ during the poling process [14, 15] modifies the result as $|F_{\vec{\Omega}_m}^{NLPC}| = 2P\cos(\delta|\Omega_m|)/m\pi$ for odd $m$ and $|F_{\vec{\Omega}_m}^{NLPC}| = 2P\sin(\delta|\Omega_m|)/m\pi$ for even $m$. But since normally $\delta \ll a_0$, this additional domain spreading can be neglected without the lost of validity.

Our calculation shows that two parallel stacked anti-stacking sequences can realize multiple photonic Bragg diffractions in 1D NLPC. In such extended para-stacking design, two anti-stacking sequences of $A = H_1 a_0$ and $B = H_2 a_0$ ($H_1 + H_2 > 2$), are para-stacked at their joint point as shown in Fig. 5. From the spatial function

$$\Gamma^{(n)}(x) = [1 - 2\text{int}(\frac{x - \text{int}(x/T)T}{A}) + 2\text{int}(\frac{x - \text{int}(x/T)T - A}{A})] [1 - 2\text{int}(\frac{x - (n-1)a_0}{a_1}) + 2\text{int}(\frac{x - (n-1)a_0 - a_1}{a_1})]$$

and defining $\Theta = A/L$, where $L = A + B$, finally we have

(i) when $m/(H_1 + H_2) = l$, where $l$ is an integer $\geq 1$

$$|F_{\vec{\Omega}_m}^{NLPC}| = \frac{2P}{\pi l}|\sin(\pi l\Phi)(2\Theta - 1)| \qquad (2)$$

(ii) when the value of $m/(H_1 + H_2)$ is not an integer

$$|F_{\vec{\Omega}_m}^{NLPC}| = \frac{2P}{\pi n}\left|\frac{\sin(\pi m\Theta)}{\sin[\pi n/(H_1+H_2)]}\right|\{\sin^2[\pi n/(H_1+H_2)]$$
$$- 2\cos[\pi n/(H_1+H_2)]\cos[\pi n(1-2\Phi)/(H_1+H_2)]\}^{1/2} \qquad (3)$$

Considering the linear feature of Eq. (2) and the oscillating feature of Eq. (3), our simulation results show that three NLPC structure factors, $|F_{\vec{\Omega}_m}^{NLPC}|$, $|F_{\vec{\Omega}_{m1}}^{NLPC}|$ and $|F_{\vec{\Omega}_{m2}}^{NLPC}|$, can be roughly equalized with large magnitude under the following conditions: $\Phi \approx 0.5$, $H_1 + H_2$ is sufficiently large and $m/(H_1 + H_2) \approx 1$, the three order numbers take successive values of $m = H_1 + H_2$, $m_1 = H_1 + H_2 + 1$, and $m_2 = H_1 + H_2 - 1$. The three equalized NLPC structure factors can simultaneously induce three photonic Bragg conditions in 1D NLPC. As an example, Fig. 6 illustrates the realization of the triple phase-matching resonances in difference frequency mixing process for wavelength-division-multiplexed (WDM) optical communication networks. Furthermore, more band signal conversion is envisaged by a simple extrapolation; for example, connecting three such extended para-stacking designs, the maximum of nine-band signal conversion can be achieved by adjusting $\Phi$, $\Theta$, and $m$.

Moving to 2D NLPC, light waves no longer propagate along one direction and the photonic Bragg condition is realized in a plane. Normally for the ferroelectric materials, the inverted domain after electric poling has circular shape [4-8]. If the shape of 2D NLPC primitive cell also possesses approximate circular symmetry, such as the square and hexagonal Bravais lattices, then $\Gamma^{(n)}(\vec{r})$ can be described by a *Circ* function. Hence, $f^{(n)}(\vec{r})$ can be written as $f^{(n)}(x,y) = +P$ for $\sqrt{x^2+y^2} \leq R_0$ and $f^{(n)}(x,y) = -P$ for $R_0 < \sqrt{x^2+y^2} \leq C_0/2$, where $R_0$ denotes the radius of the inverted circular domain and $C_0$ is the lattice constant (Fig. 1(b)). In polar coordinates, we have $F^{NLPC}_{\vec{\Omega}_{m,n}} = \frac{1}{NS_0}\sum_{n=1}^{N} 2P\int_0^{2\pi}\int_0^{R_0} d\theta r dr \exp[-i|\vec{\Omega}_{m,n}|r\cos(\theta-\psi)]$, where $S_0$ denotes the area of a 2D NLPC primitive cell. Let $\tau = \theta - \psi$, then we have

$$F^{NLPC}_{\vec{\Omega}_{m,n}} = 4P\pi \frac{R_0^2}{S_0} \frac{J_1(|\vec{\Omega}_{m,n}|R_0)}{|\vec{\Omega}_{m,n}|R_0} = 4P\Psi^{(2)}\left|\frac{J_1(|\vec{\Omega}_{m,n}|R_0)}{|\vec{\Omega}_{m,n}|R_0}\right| \qquad (4)$$

where $J_1(|\vec{\Omega}_{m,n}|R_0)$ is first-order Bessel function of the first kind. The area ratio of $4\pi R_0^2/S_0$ is defined as the 2D duty cycle $\Psi^{(2)}$. As shown in Fig. 7(a-b), except for $\vec{\Omega}_{0,1(1,0)}$, $|F^{NLPC}_{\vec{\Omega}_{m,n}}|$ equals to zero at some particular points of $\Psi^{(2)}$, causing the forbiddance of the corresponding photonic Bragg condition. Hence, whether or not a photonic Bragg diffraction can be induced in 2D NLPC is determined by both of the NLPC Ewald construction (Fig. 3) as well as the relative size of the NLPC atom.

The above study was developed as a simple analysis of the general behavior of structure factor in 2D NLPC. The results can be applied to a wide range of NLPCs with 2D Bravais lattices. It is essential to appreciate that the use of the Circ function as the mathematical presentation of the 2D NLPC shape qualifies the use of the theory. From Fig. 7, it is seen that due to the effect of Circ function, $|F_{\vec{\Omega}_{m,n}}^{NLPC}|$ does not return to zero at $\Psi^{(2)}=1$. This discrepancy from the physical reality in the square lattice becomes more evident than that in the hexagonal lattice, since the hexagonal shape looks more like a circle. However this discrepancy is not very important for small $\Psi^{(2)}$, since for the sufficiently small inversed domain, the structure factor is not expected to be very dependent on the shape of the primitive cell, and therefore Circ function can give quite good approximation. It should be noted that although 2D NLPC allows multiple photonic Bragg conditions in different directions, it is not expected for applications requiring high energy conversion efficiency. As shown in Fig. 7(c), the maximum value of $|F_{\vec{\Omega}_{m,n}}^{NLPC}|$ in 2D NLPC is always below $2P/\pi$ and hence 1D NLPC is a better choice for single parametric process since larger $|F_{\vec{\Omega}_{m}}^{NLPC}|$ is achieved and thus ensures better conversion efficiency.

In summary, a general diffractive study of the quasi-phase-matched parametric process in NLPC is presented. The modified Ewald construction is obtained, which reveals the complete set of photonic Bragg diffractions for both of the air-dielectric and dielectric-dielectric contacts. A simple model for the multiple phase-matching resonances in 1D NLPC is obtained. The parameters that affect the NLPC structure factor are investigated, which shows that although 2D NLPC permits the multiple parametric processes, 1D NLPC is a better choice than 2D NLPC for single parametric process due to the larger NLPC structure factor.

## Acknowledgements

The work is supported by A-Star of Singapore under the project No. R398000039305.

**Fig. 1.** (a) Schematic diagram of 1D NLPC with anti-stacking and para-stacking sequences. $a_0$, $a_1$, and $a_2$ denote the dimensions of NLPC primitive cell and atoms respectively. (b) Hexagonal 2D NLPC with the lattice constant of $C_0$.

**Fig. 2.** Schematic diagram of virtual wave approximation in hexagonal 2D NLPC lattice. Without regard of the energy distribution among the interacting waves, the entire parametric process in a NLPC can be treated as a propagating virtual wave with $|\xi_0|^2$ carrying the total energy of the parametric process and wave vector $\vec{K} = \Delta\vec{k}$ reflecting the phase mismatch, which makes the entire energy $|\xi_0|^2$ oscillate between the initial and the product waves.

**Fig. 3.** The NLPC Ewald construction. (a) The air-dielectric contact. C.E. is the same as $\Gamma$. (b) The dielectric-dielectric contact. C. E. deviates from $\Gamma$ by a vector of $\vec{K}_0$.

**Fig. 4.** The $|F_{\vec{\Omega}_m}^{NLPC}|$ spectra of the anti-stacking and para-stacking sequences. According to Refs [1, 4], the reciprocal vector is defined as $|\vec{\Omega}_m| = m2\pi/a_0$ for anti-stacking sequence and $|\vec{\Omega}_m| = m\pi/a_0$ for para-stacking sequence.

**Fig. 5.** Schematic diagram of the extended para-stacking design.

**Fig. 6.** Triple-band signal conversion for WDM networks. The center wavelengths of the pump, input signal, and converted output waves are chosen to be 775nm, 1545nm, and 1555nm to fit the optical communication window. In LiNbO$_3$, the lattice constant $a_0$ is calculated to be 16.92μm from the Sellmeier equation. As indicated by the black circles, near equalization of the three NLPC structure factors is achieved at $\Theta = 0.35$ or (1-0.35). The equalization points will increase as the value of $H_1 + H_2$ increases.

**Fig. 7.** (a-b) 2D NLPC structure factor. $m$ and $n$ are mutual to each other and $|\vec{\Omega}_{m,n}| \equiv |\vec{\Omega}_{n,m}|$. (c) Comparison of the NLPC structure factors in 1D and 2D NLPCs.

**Fig. 1**

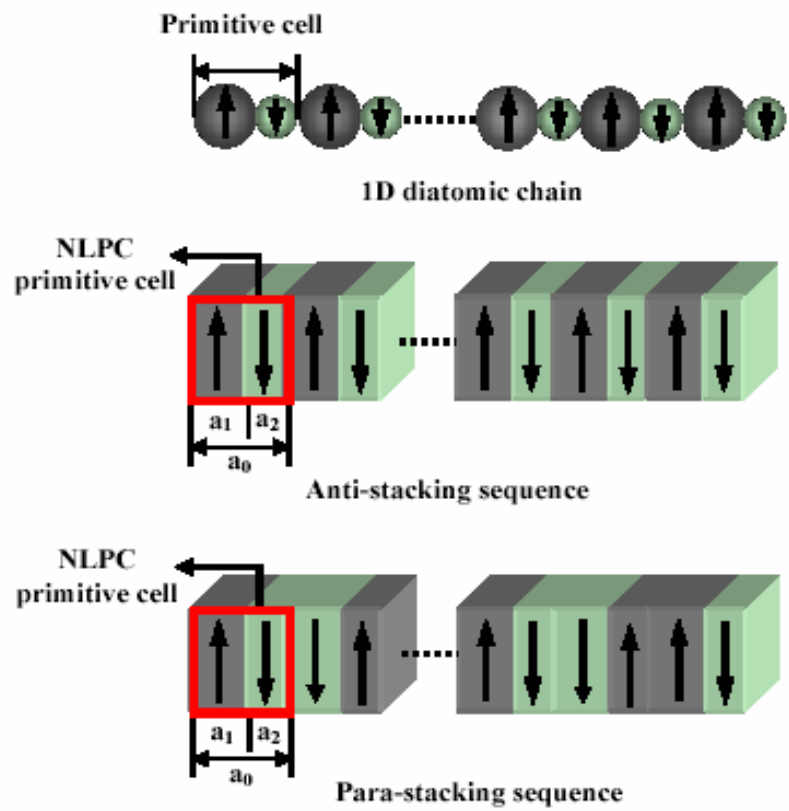

Fig. 1(a)

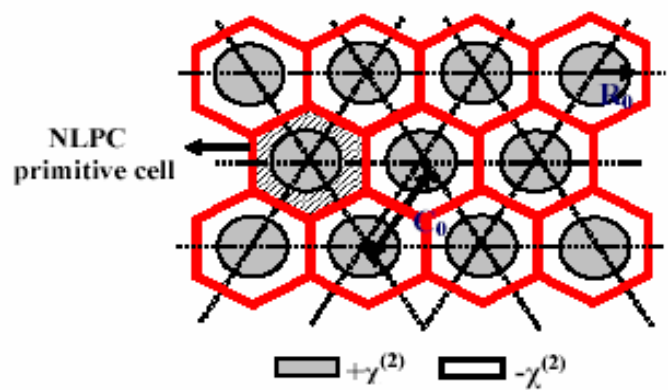

Fig. 1(b)

**Fig. 2**

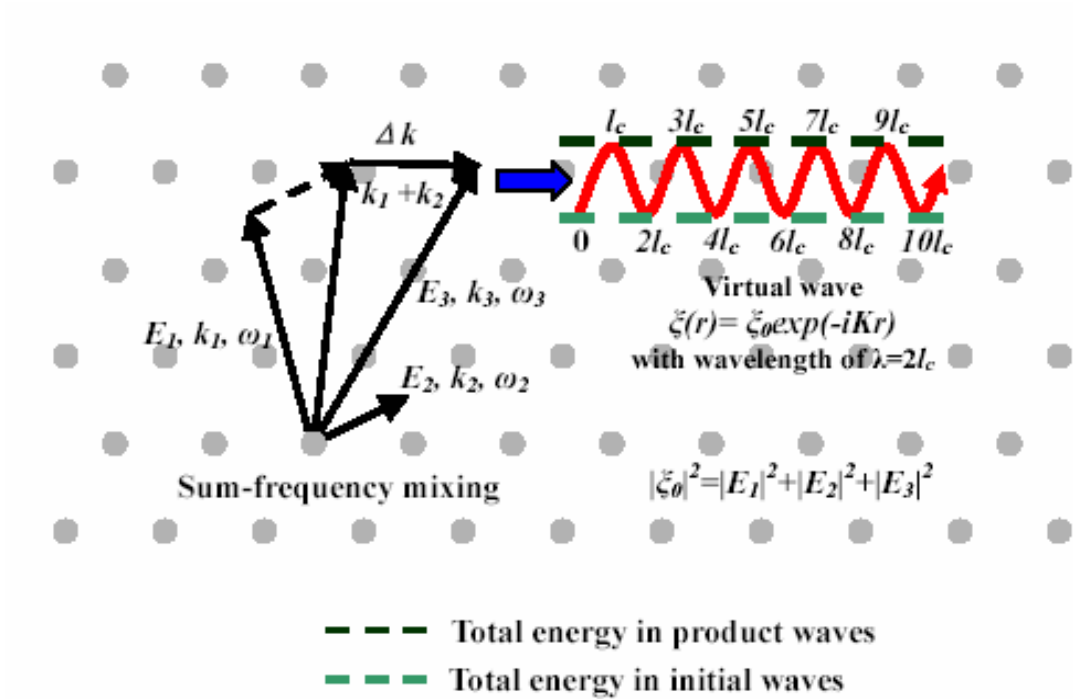

- - - Total energy in product waves
- - - Total energy in initial waves

**Fig. 3**

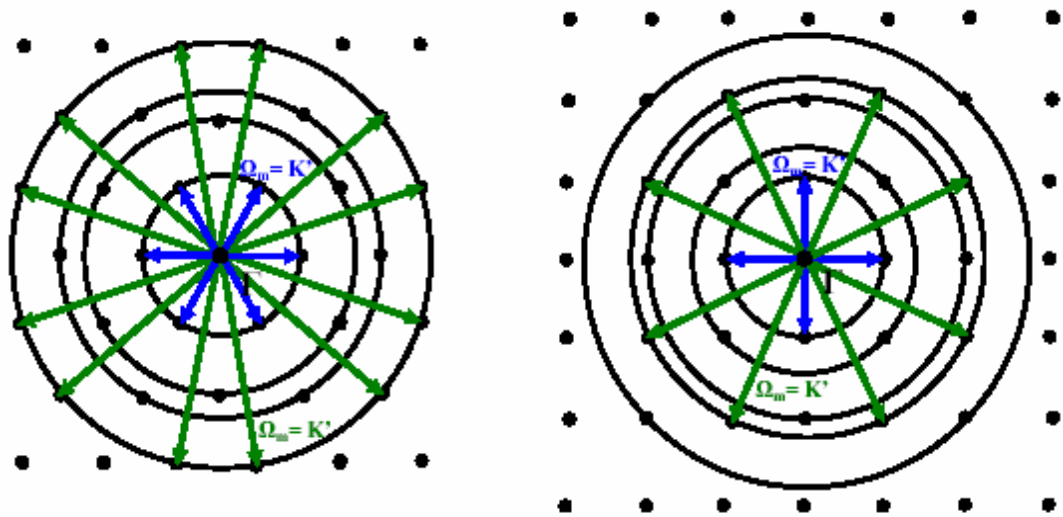

Fig. 3(a)

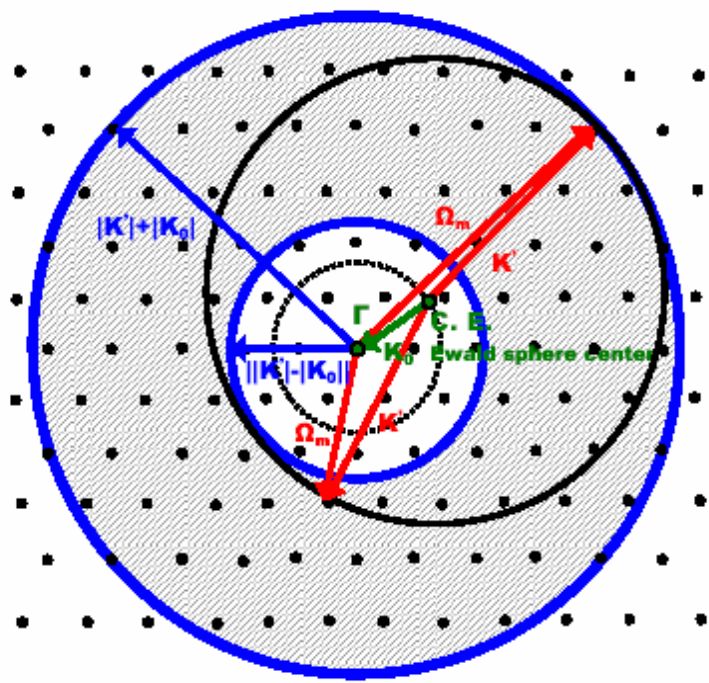

Fig. 3(b)

**Fig. 4**

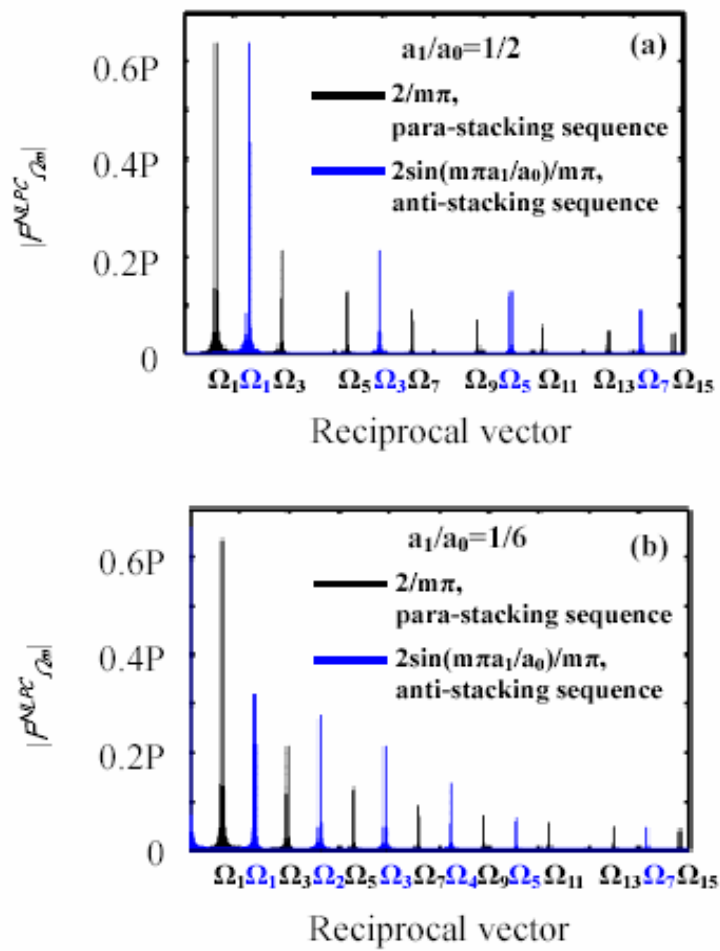

**Fig. 5**

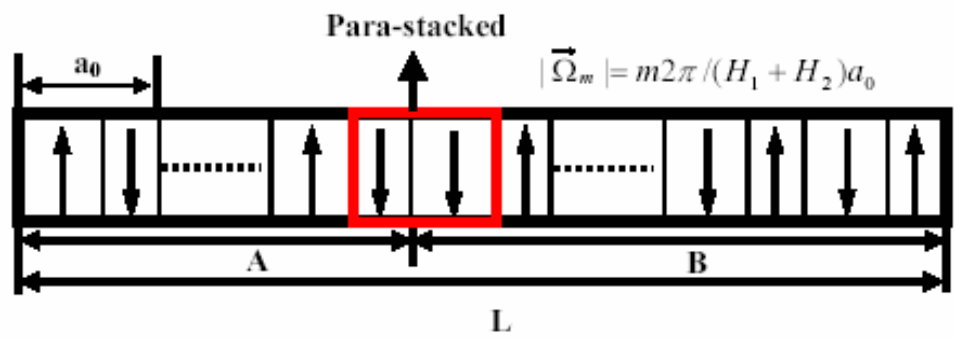

**Fig.6**

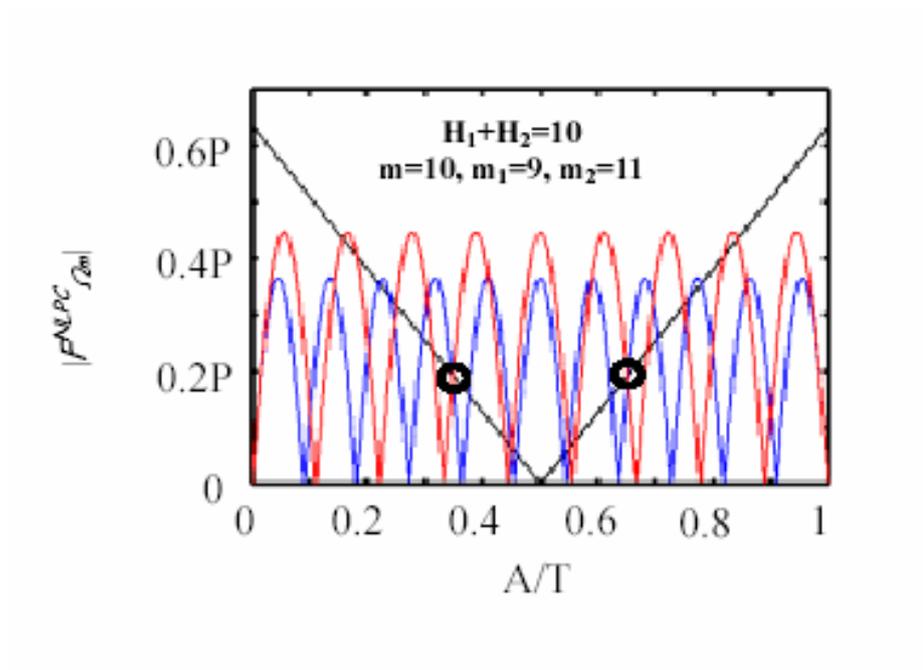

**Fig. 7**

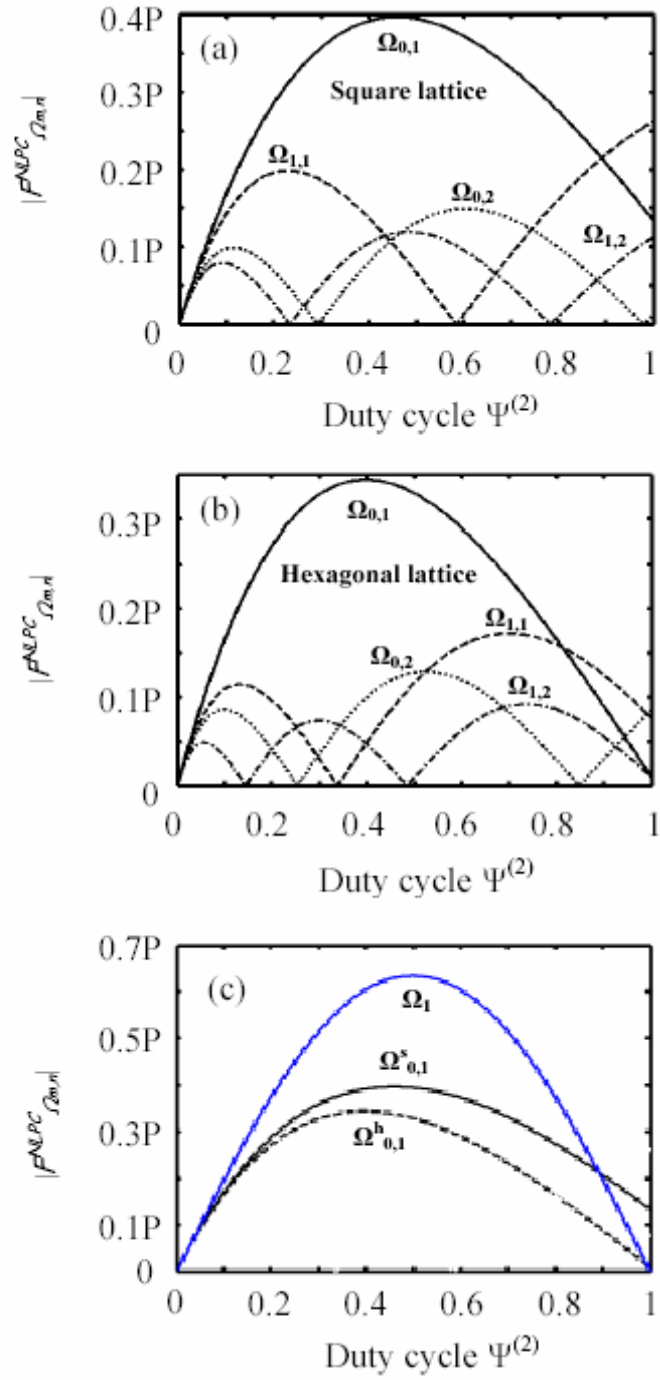